\documentclass[
    ,final            
  ]
  {aipproc}

\layoutstyle{6x9}
\usepackage{amssymb}
\usepackage{color}

\begin{document}

\title[]{The moving hotspot model for kHz QPOs in accreting neutron stars.}


\author{Matteo Bachetti}{
  address={Centre d'Etude Spatiale des Rayonnements, Toulouse, France}
}

\author{Marina M. Romanova}{
  address={Cornell University, Ithaca NY, USA}
}

\begin{abstract}
3D MHD simulation of accretion onto neutron stars have shown in the
last few years that the footprint (hotspot) of the accretion flow
changes with time. Two different kinds of accretion, namely the funnel
flow and the equatorial accretion produced by instabilities at the
inner disk, produce different kinds of motion of the hotspot. The
funnel flow produces hotspots that move around the magnetic pole,
while instabilities produce other hotspots that appear randomly and
move along the equator or slightly above. The angular velocities of
the two hotspots are different, the equatorial one being higher and
both close to the Keplerian velocity in the inner region.
Modeling of the lightcurves of these hotspots with Monte Carlo
simulations show that the signatures produced in power specra by them,
if observed, are QPOs plus low frequency components. Their
frequencies, general behavior and features describe correctly most of
the properties of kHz QPOs, if we assume the funnel flow hotspots as
the origin of the lower kHz QPO and instabilities as the origin of the
upper kHz QPO.
\end{abstract}

\maketitle

\section{The kHz QPO phenomenon}
The lightcurves of X-ray binaries often show a number of variability phenomena, that populate the power spectra of their lightcurves with different kinds of noise and oscillatory components. They are normally fitted with Lorentzian curves \citep{Belloni:2002p8449} and classified in terms of their characteristic frequency $\nu$ and their width $\Delta\nu$. Depending on the quality factor $Q=\nu/\Delta\nu$, features are classified as broadband noise ($Q\lesssim 2$), quasi-periodic oscillations (QPO, $Q \gtrsim 2$), coherent pulsations ($\Delta\nu$ smaller than the frequency bin). Many weakly-magnetized accreting neutron stars show a particular kind of QPOs, whose frequencies range from $300$ to $1300$~Hz, called for this reason kHz~QPOs \citep{vanderKlis:1996p2322}. Very often kHz~QPOs appear in pairs, with the two peaks, labelled ``upper'' and ``lower'', at frequencies differing by about $300$~Hz. Lower and upper peak generally differ for their properties, the lower peak being normally more coherent and appearing in a shorter frequency range \citep{diSalvo:2003p5128,Barret:2006p7645}. Their frequencies are correlated with the ones of lower-frequency phenomena. KHz QPOs tend to have higher frequencies at higher countrates, but the correlation is not straightforward. On small timescales, in fact, the variation of the frequency with the countrate is usually very rapid, with the frequency going up of hundreds of Hz with slight increases of the countrate. If one plots the frequency of the oscillations versus the countrate, it is evident that the variation happens on different tracks, almost parallel to each other but distinct. Instead, there is a good correlation between the frequency and the spectral state \citep{Mendez:1999p4106}, and at the lower and upper limits of their range the frequency difference between the peaks decreases \citep[e.g.]{Boutloukos:2006p5053}. 
Possible models for kHz QPOs include a number of different phenomena, including but not limited to relativistic resonances \citep[e.g.][]{Kluzniak:2001p4749}, Lense-Thirring relativistic precession of matter in preferred orbits \citep[e.g.][]{Stella:1998p4748}, magnetic field-driven oscillations \citep[e.g.][]{Li:2004p9308}. 

\section{Moving hot spots and QPOs}
\begin{figure}
\centering
$ \begin{array}{cc}
 \includegraphics[width=2in]{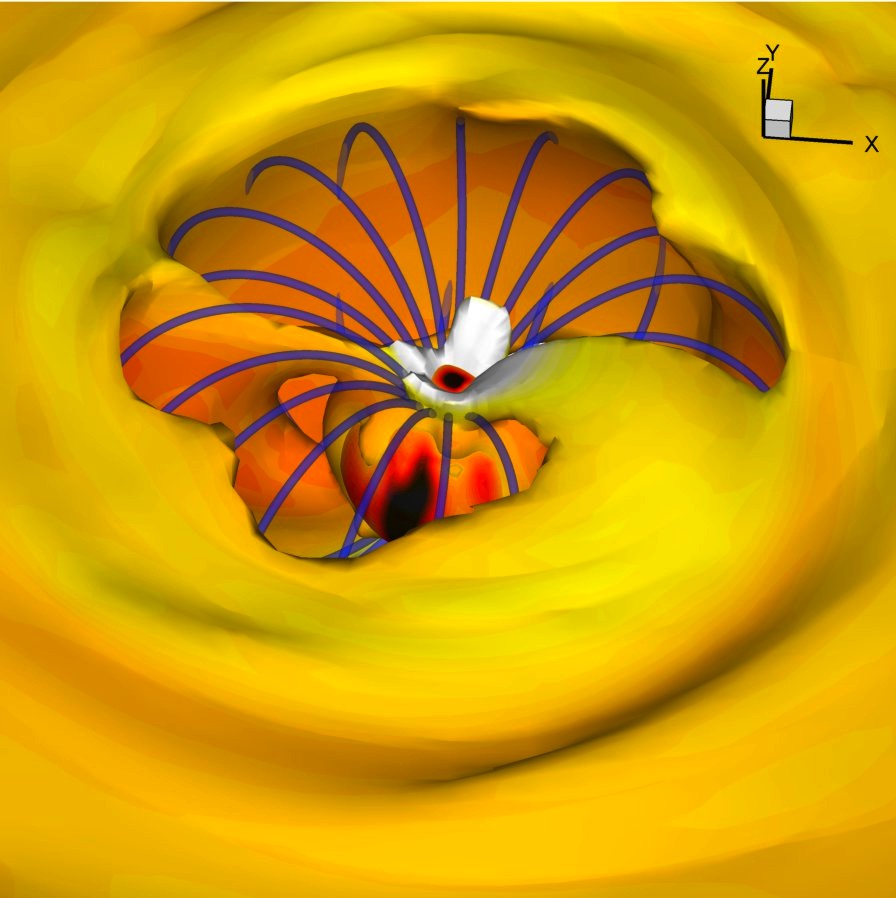} &
 \includegraphics[width=3in]{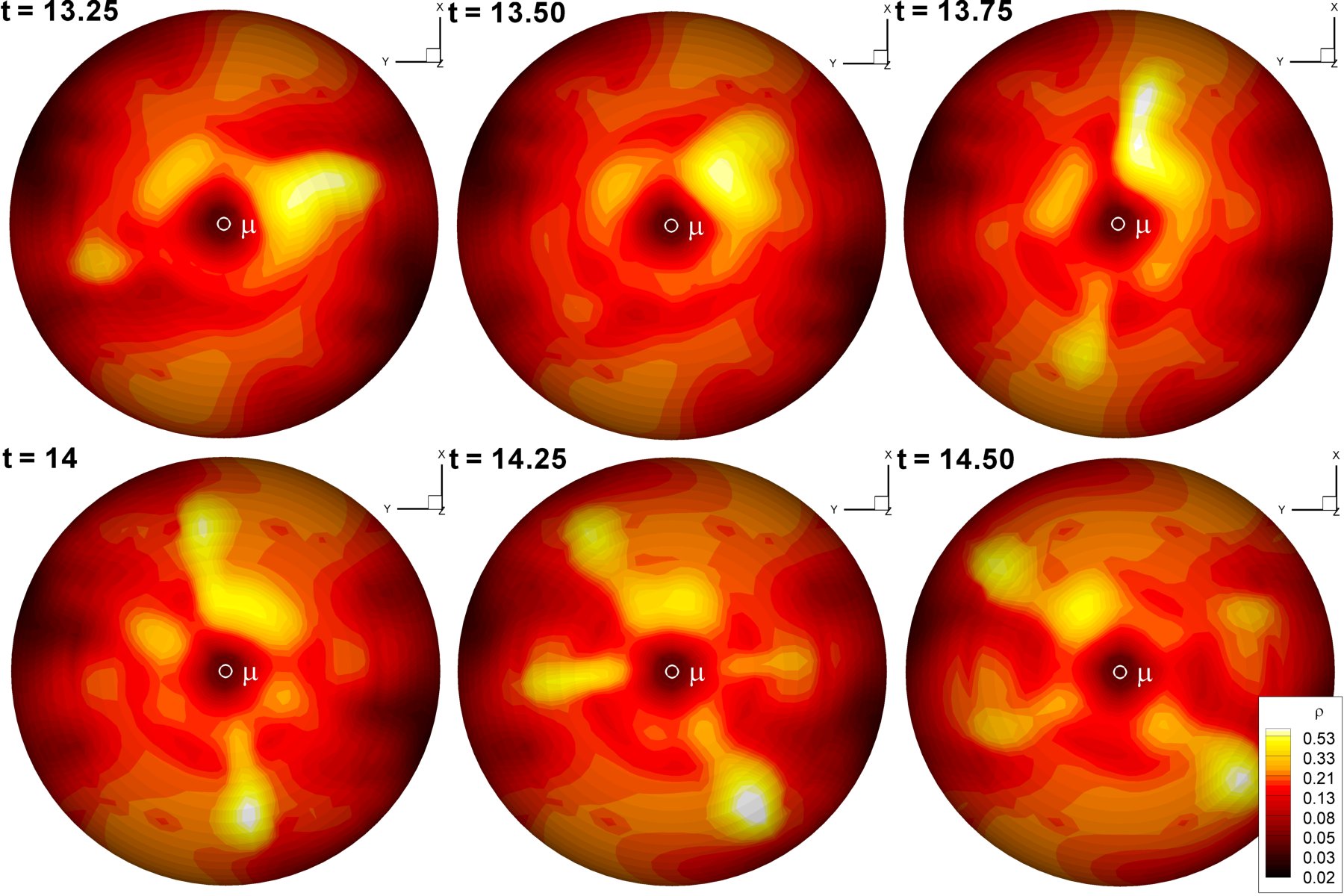}
 \end{array} $
\caption{In this figure it is possible to observe ({\em left}) the matter streams falling on the polar region from the funnel flow and the ones from instabilities at the inner radius and ({\em right}) the hotspots left on the surface by these two different accretion mechanisms \citep{Bachetti:2010p7648}}
\label{fig:3dspots}
\end{figure}
Global 3D magnetohydrodynamical simulations of accreting, weakly-magnetized neutron stars show (e.g. \cite{Romanova:2004p4095}, \cite{Kulkarni:2009p5067},\cite{Romanova:2009p6603}) that most accretion configurations produce hotspots on the surface of the star. These spots are confined on a small region of the star (hence, by all means, ``fixed'') only in one case: when accretion is on the polar region through funnel flow and the misalignment angle between the magnetic field and the rotation axis is sufficiently large. In this case, in fact, matter finds a ``preferred path'' during its fall, that takes near the magnetic pole, fixed in the star's reference frame. This hotspot is thus tied to a region of the star and rotates with it, and an observer at infinity will see the light coming from the hotspot modulated at the rotational frequency of the star, similarly to what is observed in x-ray pulsars. In all other cases, namely funnel flow with small misalignment angle and unstable accretion in the equatorial region, the hotspots tend to move (Fig.~\ref{fig:3dspots}): in fact, in these cases the magnetic field does not change considerably along $\phi$ and thus matter is able to fall in different places of the surface, or in different positions around the magnetic pole. If a stream of matter is formed that falls on the surface, the corresponding hotspot will move with an angular velocity similar to the velocity at the starting point of the stream in the disk.

If we model with Monte Carlo techniques these moving hotspots, we extract a lightcurve and we take a Fourier Transform of it, we immediately find some interesting things. First of all, the shape of the peaks at the rotational frequency of the hotspot is a ``bump'', a QPO. This is because, even if the rotational frequency of the hotspots is fixed, their duration is limited and thus Fourier theory shows that the Fourier transform is not a single-frequency peak but a bell-shaped function (in the ideal case, a $\mathrm{sinc}$ function). Moreover, the rotational frequency of the hotspots is not perfectly stable, and thus the broadness of the peak increases. Second, the possible frequencies to observe at a time are two: one given by the hotspots moving around the magnetic pole, one given by the hotspots produced through instabilities. The frequency of the instability hotspots is that of the inner edge of the disk, not very stable, and the instabilities have a relatively short duration. The polar region hotspots, instead, are more stable and last longer. Therefore, the features they produce in power spectra are generally more coherent than those from instabilities. The frequency is lower because the accretion on the polar region does not come from the very edge of the disk, but from a larger region behind it. Finally, the moving hotspots produce red noise, whose shape and amplitude depend on the statistical distribution of amplitude and intensity of the hotspots. The striking similarities between the behavior of these two different ways to produce hotspots and the two kHz QPOs lead us to propose this mechanism as a possible explanation for the kHz QPO phenomenon.

\section{Testing the reliability of the model}
\begin{figure}
\centering
$ \begin{array}{cc}
 \includegraphics[width=2.5in]{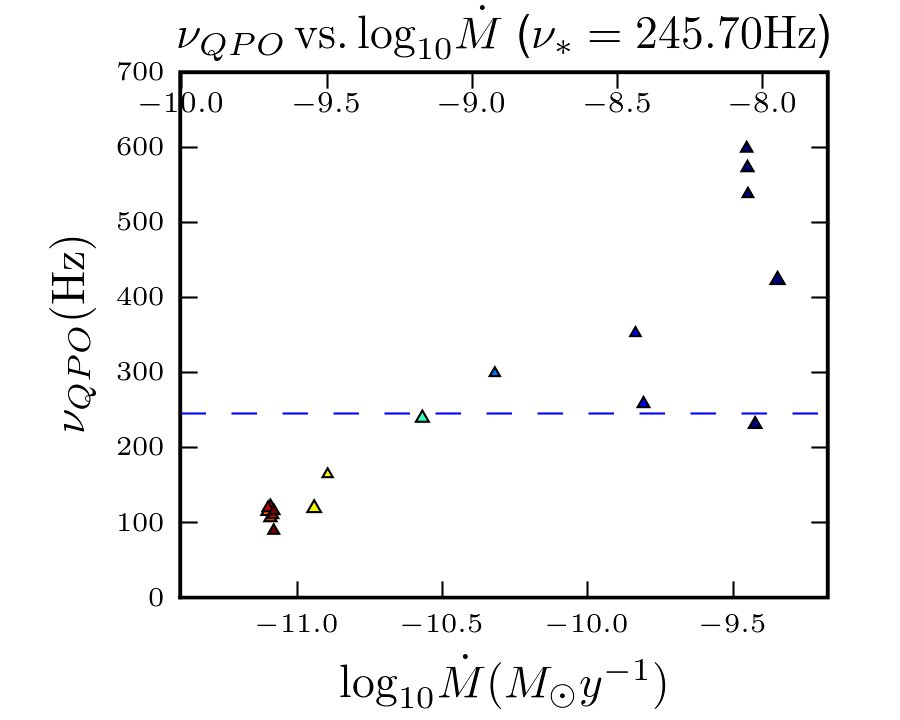} &
 \includegraphics[width=2.5in]{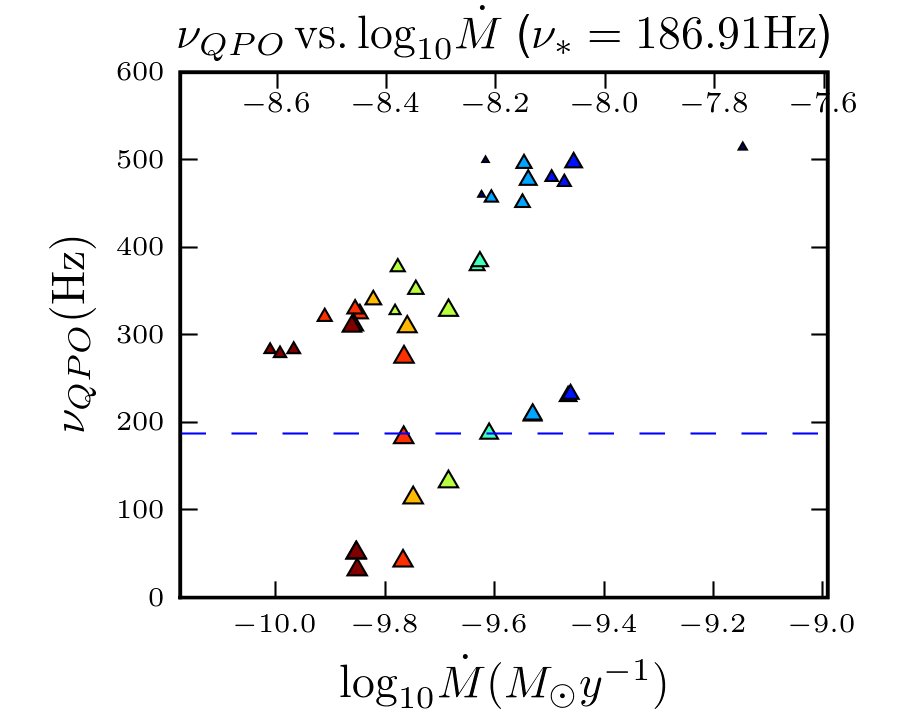}
 \end{array} $
\caption{Frequency variation of the hot spot movements with the mass accretion rate \cite{Bachetti:2010p7648}. Results are scaled for the two cases $B=10^8$G and $B=5\cdot10^8$G. The first case (right) shows only stable accretion from the funnel flow. The hot spot moves around the magnetic pole with a measurable angular velocity, which we plot against the mass accretion rate $\dot{M}$.
The second case (left) shows instabilities at the inner radius. In this case we have both accretion onto the magnetic pole with the usual funnel flow, like in the first case, and accretion from instabilities, moving approximately at the frequency of the inner disk. Hence the ``double track'' in the plot.}
\label{fig:numdot}
\end{figure}

To propose this mechanism as a viable way to explain QPO, we needed to be sure that the frequency of rotation of the hotspots followed a consistent behavior. First of all, the hotspots had to show up in different simulations with similar parameters in a similar way. Then, their frequencies had to change in a predictable way.

For this reason, we performed a number of simulations differing only by the mass accretion rate. All the systems reproduced in the simulations have $M=1.4M_{\odot}$, $R=10$km and misaligment angle $\theta=2^{\circ}$. One system has a rotational period $\tau=4.1$ms, the other $\tau=5.4$ms. In Fig.~\ref{fig:numdot} we show the frequencies of all hotspots appearing on the surface, plotted against the mass accretion rate $\dot{M}$. In one case the accretion is only through the funnel flow, in the other both through the funnel flow and the instabilities. 

As we can see from the figure, the frequencies of the hotspots in the funnel flow and from the instabilities increase with frequency, forming two distinct tracks in the $\nu-\dot{M}$ plane. Their frequency difference is around 200~Hz. Their consistent behavior tells us that these phenomena are something intrinsic to the accretion process as modeled in MHD simulations, and most likely to the accretion in real systems. 

\section{Conclusions}
The results here reported represent a viable explanation for the kHz QPO phenomenon. To summarize the main points: $1$) MHD simulations show that multiple, moving hotspots are formed on the surface of neutron stars, both in the funnel flow (if the misalignment angle is small enough) and from instabilities; $2$) the movement of these hotspots produces a modulation of the lightcurve, whose appearance on a power spectrum is a QPO plus red noise components; $3$) the two different accretion mechanisms produce qualitatively different QPOs, the ones from the funnel flow being usually more coherent and at a lower frequency; $4$) their frequency separation is consistent with the frequency separation of real-life kHz QPOs.
For more information, see \citep{Bachetti:2010p7648}.


\begin{theacknowledgments}
Research supported by INFN (MB) and the contract ASI-INAF I/088/06/0 for High Energy Astrophysics (MB).
MMR was supported in part by NASA grant NNX08AH25G and by NSF grant AST-0807129. Simulations performed on NASA (Pleiades, Columbia), and CINECA (BCX, SP6, under the 2008-2010 INAF/CINECA agreement) supercomputers. 
\end{theacknowledgments}



\bibliographystyle{aipproc}   


\begin{thebibliography}{14}
\expandafter\ifx\csname natexlab\endcsname\relax\def\natexlab#1{#1}\fi
\providecommand{\enquote}[1]{``#1''}
\expandafter\ifx\csname url\endcsname\relax
  \def\url#1{\texttt{#1}}\fi
\expandafter\ifx\csname urlprefix\endcsname\relax\def\urlprefix{URL }\fi
\providecommand{\eprint}[2][]{\url{#2}}

\bibitem[Belloni et~al.(2002)]{Belloni:2002p8449}
T.~Belloni, D.~Psaltis, and M.~van~der Klis, \emph{ApJ}
  \textbf{572}, 392 (2002).

\bibitem[van~der Klis et~al.(1996)]{vanderKlis:1996p2322}
M.~van~der Klis, J.~H. Swank, W.~Zhang, K.~Jahoda, E.~Morgan, W.~H.~G. Lewin,
  B.~Vaughan, and J.~van Paradijis, \emph{ApJ} \textbf{469},
  L1 (1996).

\bibitem[di~Salvo et~al.(2003)]{diSalvo:2003p5128}
T.~di~Salvo, M.~M{\'e}ndez, and M.~van~der Klis, \emph{A\&A} \textbf{406}, 177 (2003).

\bibitem[Barret et~al.(2006)]{Barret:2006p7645}
D.~Barret, J.-F. Olive, and M.~C. Miller, \emph{MNRAS} \textbf{370}, 1140 (2006).

\bibitem[M{\'e}ndez et~al.(1999)]{Mendez:1999p4106}
M.~M{\'e}ndez, M.~van~der Klis, E.~C. Ford, R.~A.~D. Wijnands, and J.~van
  Paradijis, \emph{ApJ} \textbf{511}, L49 (1999).

\bibitem[Boutloukos et~al.(2006)]{Boutloukos:2006p5053}
S.~Boutloukos, M.~van~der Klis, D.~Altamirano, M.~Klein-Wolt, R.~A.~D.
  Wijnands, P.~G. Jonker, and R.~P. Fender, \emph{ApJ}
  \textbf{653}, 1435 (2006).

\bibitem[Kluzniak and Abramowicz(2001)]{Kluzniak:2001p4749}
W.~Kluzniak, and M.~A. Abramowicz, \emph{eprint arXiv} p. 5057 (2001).

\bibitem[Stella and Vietri(1998)]{Stella:1998p4748}
L.~Stella, and M.~Vietri, \emph{ApJL}
  \textbf{492}, L59 (1998).

\bibitem[Li and Narayan(2004)]{Li:2004p9308}
L.-X. Li, and R.~Narayan, \emph{ApJ} \textbf{601}, 414
  (2004).

\bibitem[Bachetti et~al.(2010)]{Bachetti:2010p7648}
M.~Bachetti, M.~M. Romanova, A.~Kulkarni, L.~Burderi, and T.~di~Salvo,
  \emph{MNRAS} \textbf{403}, 1193
  (2010).

\bibitem[Romanova et~al.(2004)]{Romanova:2004p4095}
M.~M. Romanova, G.~V. Ustyugova, A.~V. Koldoba, and R.~V.~E. Lovelace,
  \emph{ApJ} \textbf{610}, 920 (2004).

\bibitem[Kulkarni and Romanova(2009)]{Kulkarni:2009p5067}
A.~K. Kulkarni, and M.~M. Romanova, \emph{MNRAS} \textbf{398}, 701 (2009).

\bibitem[Romanova and Kulkarni(2009)]{Romanova:2009p6603}
M.~M. Romanova, and A.~K. Kulkarni, \emph{MNRAS} \textbf{398}, 1105 (2009).

\bibitem[Menna et~al.(2003)]{Menna:2003p1045}
M.~T. Menna, L.~Burderi, L.~Stella, N.~Robba, and M.~van~der Klis, \emph{arXiv}
  \textbf{astro-ph} (2003),
\eprint{astro-ph/0303606v1}.

\end{thebibliography}

\IfFileExists{\jobname.bbl}{}
 {\typeout{}
  \typeout{******************************************}
  \typeout{** Please run "bibtex \jobname" to optain}
  \typeout{** the bibliography and then re-run LaTeX}
  \typeout{** twice to fix the references!}
  \typeout{******************************************}
  \typeout{}
 }

\end{document}